\documentstyle[12pt]{article}
\newlength{\pubnumber} 

\catcode`\@=11
\@addtoreset{equation}{section}
\def\section{\@startsection{section}{1}{\z@}{3.5ex plus 1ex minus .2ex}
 {2.3ex plus .2ex}{\large\bf}}
\def\subsection{\@startsection{subsection}{2}{\z@}{2.3ex plus .2ex}
 {2.3ex plus .2ex}{\bf}}

\begin{document}

\begin{titlepage}
\samepage{
\setcounter{page}{1}
\rightline{CERN--TH--97/237}
\rightline{UFIFT-HEP-97--24}
\rightline{CPT--TAMU--36/97}
\rightline{ACT--14/97}
\rightline{\tt hep-th/9709049}
\rightline{September 1997}
\vfill
\begin{center}
 {\Large \bf M-Theory Model-Building and Proton Stability\\}
\vfill
\vspace{.25in}
 {\large John Ellis$^{1}$,
   Alon E. Faraggi$^{2}$,
   $\,$and$\,$ D.V. Nanopoulos$^3$\\}
\vspace{.25in}
 {\it  $^{1}$ Theory Division, CERN, CH-1211, Geneva, Switzerland \\}
\vspace{.05in}
 {\it  $^{2}$ Institute for Fundamental Theory, Department of Physics,\\
              University of Florida, Gainesville, FL  32611, USA\\}
\vspace{.05in}
 {\it  $^{3}$ Center for
Theoretical Physics, Dept. of Physics,
Texas A \& M University, College Station, TX 77843-4242, USA, and \\
Astroparticle Physics Group, Houston
Advanced Research Center (HARC), The Mitchell Campus,
Woodlands, TX 77381, USA, and \\
Academy of Athens, Chair of Theoretical Physics,
Division of Natural Sciences, 28 Panepistimiou Avenue,
Athens 10679, Greece.\\}
\end{center}
\vfill
\begin{abstract}
  {\rm
We study the problem of baryon stability in M theory, starting
from realistic four-dimensional string models constructed using
the free-fermion formulation of the weakly-coupled heterotic string.
Suitable variants of these models manifest an enhanced custodial
gauge symmetry that forbids to all orders the appearance of dangerous
dimension-five baryon-decay operators. We exhibit the underlying
geometric (bosonic) interpretation of these models, which have a
$Z_2 \times Z_2$ orbifold structure similar, but not identical, to
the class of Calabi-Yau threefold compactifications of M and F theory
investigated by Voisin and Borcea. A related generalization of their
work may provide a solution to the problem of proton stability in
M theory.}
\end{abstract}
\vfill
\smallskip}
\end{titlepage}

\setcounter{footnote}{0}

\def\beq{\begin{equation}}
\def\eeq{\end{equation}}
\def\beqn{\begin{eqnarray}}
\def\eeqn{\end{eqnarray}}
\def\Tr{{\rm Tr}\,}
\def\KM{{Ka\v{c}-Moody}}

\def\ie{{\it i.e.}}
\def\etc{{\it etc}}
\def\eg{{\it e.g.}}
\def\half{{\textstyle{1\over 2}}}
\def\third{{\textstyle {1\over3}}}
\def\quarter{{\textstyle {1\over4}}}
\def\m{{\tt -}}
\def\p{{\tt +}}

\def\rep#1{{\bf {#1}}}
\def\slash#1{#1\hskip-6pt/\hskip6pt}
\def\slk{\slash{k}}
\def\GeV{\,{\rm GeV}}
\def\TeV{\,{\rm TeV}}
\def\y{\,{\rm y}}
\def\SM{Standard-Model }
\def\SUSY{supersymmetry }
\def\SSM{supersymmetric standard model}
\def\vev#1{\left\langle #1\right\rangle}
\def\l{\langle}
\def\r{\rangle}

\def\Htw{{\tilde H}}
\def\chibar{{\overline{\chi}}}
\def\qbar{{\overline{q}}}
\def\ibar{{\overline{\imath}}}
\def\jbar{{\overline{\jmath}}}
\def\Hbar{{\overline{H}}}
\def\Qbar{{\overline{Q}}}
\def\abar{{\overline{a}}}
\def\alphabar{{\overline{\alpha}}}
\def\betabar{{\overline{\beta}}}
\def\tautwo{{ \tau_2 }}
\def\calF{{\cal F}}
\def\calP{{\cal P}}
\def\calN{{\cal N}}
\def\smallmatrix#1#2#3#4{{ {{#1}~{#2}\choose{#3}~{#4}} }}
\def\bone{{\bf 1}}
\def\V{{\bf V}}
\def\b{{\bf b}}
\def\N{{\bf N}}
\def\bQ{{\bf Q}}
\def\t#1#2{{ \Theta\left\lbrack \matrix{ {#1}\cr {#2}\cr }\right\rbrack }}
\def\C#1#2{{ C\left\lbrack \matrix{ {#1}\cr {#2}\cr }\right\rbrack }}
\def\tp#1#2{{ \Theta'\left\lbrack \matrix{ {#1}\cr {#2}\cr }\right\rbrack }}
\def\tpp#1#2{{ \Theta''\left\lbrack \matrix{ {#1}\cr {#2}\cr }\right\rbrack }}


\def\inbar{\,\vrule height1.5ex width.4pt depth0pt}

\def\IC{\relax\hbox{$\inbar\kern-.3em{\rm C}$}}
\def\IQ{\relax\hbox{$\inbar\kern-.3em{\rm Q}$}}
\def\IR{\relax{\rm I\kern-.18em R}}
 \font\cmss=cmss10 \font\cmsss=cmss10 at 7pt
\def\IZ{\relax\ifmmode\mathchoice
 {\hbox{\cmss Z\kern-.4em Z}}{\hbox{\cmss Z\kern-.4em Z}}
 {\lower.9pt\hbox{\cmsss Z\kern-.4em Z}}
 {\lower1.2pt\hbox{\cmsss Z\kern-.4em Z}}\else{\cmss Z\kern-.4em Z}\fi}

\def\AEF{A.E. Faraggi}
\def\KRD{K.R. Dienes}
\def\JMR{J. March-Russell}
\def\NPB#1#2#3{{\it Nucl.\ Phys.}\/ {\bf B#1} (19#2) #3}
\def\PLB#1#2#3{{\it Phys.\ Lett.}\/ {\bf B#1} (19#2) #3}
\def\PRD#1#2#3{{\it Phys.\ Rev.}\/ {\bf D#1} (19#2) #3}
\def\PRL#1#2#3{{\it Phys.\ Rev.\ Lett.}\/ {\bf #1} (19#2) #3}
\def\PRT#1#2#3{{\it Phys.\ Rep.}\/ {\bf#1} (19#2) #3}
\def\MODA#1#2#3{{\it Mod.\ Phys.\ Lett.}\/ {\bf A#1} (19#2) #3}
\def\IJMP#1#2#3{{\it Int.\ J.\ Mod.\ Phys.}\/ {\bf A#1} (19#2) #3}
\def\nuvc#1#2#3{{\it Nuovo Cimento}\/ {\bf #1A} (#2) #3}
\def\etal{{\it et al,\/}\ }

\hyphenation{su-per-sym-met-ric non-su-per-sym-met-ric}
\hyphenation{space-time-super-sym-met-ric}
\hyphenation{mod-u-lar mod-u-lar--in-var-i-ant}


\setcounter{footnote}{0}

There has recently been important progress
towards a better understanding of the
underlying non--perturbative formulation
of superstring theory. The picture which emerges is that all the superstring
theories in ten dimensions, as well as 11-dimensional supergravity, which
were previously thought to be distinct,
are in fact different limits of a single fundamental theory,
often referred to as M (or F) theory \cite{reviewsmf}.
To elevate this new mathematical understanding
into contact with experimentally-oriented
physics is a rewarding challenge.
Different directions may be followed in pursuing
this endeavour. On the one hand,
one may look {\it ab initio} for generic phenomenological properties
which may characterize the fundamental M (or F) theory. Or, on
the other hand, one may adapt the technologies that have
been developed for the analysis of realistic classes of heterotic string
solutions
in four dimensions~\cite{ffmreviews}, and explore the extent to which they
may apply
in the context of M and F theory.

Following the first line of thought, one of the
issues in superstring phenomenology
that received
a great deal of attention is the problem of superstring
gauge coupling unification. As is well known,
the discrepancy between the Grand Unification scale
of around $2 \times 10^{16}$ GeV estimated by extrapolating
naively from the measurements at LEP and elsewhere \cite{mssmcop}
and the estimate of around $4 \times 10^{17}$ GeV
found in weakly-coupled heterotic string theory \cite{scop} may be
removed if the Theory of Everything is
M theory in a strong-coupling limit, corresponding
to an eleventh dimension that is considerably larger
than the naive Planck length \cite{witten}. This scenario would
explain naturally why the value of sin$^2 \theta_W$
measured at accelerators is in good agreement with
minimal supersymmetric GUT predictions, a feature
not shared by generic string models with extra particles at intermediate
scales, large Planck-scale threshold corrections, or
different Kac-Moody levels for different gauge group factors \cite{df}.

This economic strong-coupling solution to the reconciliation
of the minimal supersymmetric GUT and string unification scales
may be desirable
for resolving other phenomenological issues,
such as stabilizing the dilaton vacuum expectation value and
selecting the appropriate vacuum point in moduli space~\cite{Madv}.
However, as is only too often the case,
closing the door for one genie may open a
door for another. In this case, we fear
that the problems associated with proton
stability will resurface and in fact worsen.

This has been a prospective problem for a quantum theory
of gravity ever since the no-hair theorems were discovered
and it was realized that non-perturbative vacuum
fluctuations could engender baryon decay~\cite{Hawking}, in the absence
of any custodial exact (gauge) symmetry.
This problem became particularly acute with the
advent of supersymmetric GUTs, when it was realized
that effective dimension-five operators of the form
\beq
Q Q Q L\label{d5operator}
\eeq
could induce rapid baryon decay. Operators of this form could be
generated either by the exchange of GUT particles~\cite{WSY} or by quantum
gravity
effects~\cite{EHNT}. Specifically, in the context of string theory, such
an operator could in general be induced by
the exchange of heavy string modes. In this case, the coefficient of
the operator (\ref{d5operator}) would be suppressed by one inverse
power of the effective string scale $M$.
In the perturbative heterotic string solutions studied heretofore,
this scale is of the order $M\sim10^{18}$ GeV,
whilst in the proposed non--perturbative M-theory solution
to the string-scale gauge-coupling unification problem
this scale would be of the order $M\sim10^{16}$ GeV.
Thus, the magnitude
of the effective dimension-five operator (\ref{d5operator})
may increase by $\sim$
two orders of magnitude. As proton stability considerations
severely restrict the magnitude of such operators,
and as the general expectation is that this kind of operator
is abundant in a generic superstring vacuum,
it would seem that the M-theory resolution of the
problem of string-scale gauge-coupling unification,
we have re--introduced a far more serious problem,
namely that of baryon decay.

A full M- (or F-)theory solution to this problem lies beyond
technical reach at this time. However, we believe that
useful insight into this problem may be obtained by
examining perturbative heterotic string models in four
dimensions that possess some realistic
properties, identifying symmetries that guarantee the
absence of dangerous dimension-five operators to all orders
in string perturbation theory, and then investigating the
possibility of elevating such models to a full non-perturbative
M- (or F-)theory formulation.

For this purpose, we choose to investigate
the three-generation superstring models
\cite{fsu5,fny,alr,eu,cus,lykken}
derived in the free-fermion formulation \cite{fff}. This
construction produces a large
number of three-generation models with
different phenomenological characteristics,
some of which are especially appealing.
This class of models corresponds to $Z_2\times Z_2$
orbifold compactification at the maximally-symmetric
point in the Narain moduli space~\cite{Narain}. The emergence
of three generations is correlated with the
underlying $Z_2\times Z_2$ orbifold structure.
Detailed studies of specific models have
revealed that these models may
explain the qualitative structure of the
fermion mass spectrum~\cite{spectrum} and could form the basis of
a realistic superstring model.
We refer the interested reader to several
review articles which summarize the phenomenological
studies of this class of models \cite{ffmreviews}.

For our purposes here, let us recall the main structures
underlying this class of models.
In the free-fermion formulation~\cite{fff}, a model
is defined by a set of boundary condition basis
vectors, together with the related one--loop
GSO projection coefficients, that are constrained
by the string consistency constraints. The basis vectors,
$b_k$, span a finite additive group, $\Xi=\sum_k n_kb_k$
where $n_k=0,\cdots,N_{z_k}-1$.
The physical states in the Hilbert space of
a given sector $\alpha\in\Xi$, are obtained
by acting on the vacuum with bosonic and fermionic
operators and by applying the generalized GSO
projections. The $U(1)$ charges $Q(f)$ corresponding
to the unbroken Cartan generators of the four-dimensional gauge group,
which are in one-to-one correspondence
with the $U(1)$ currents ${f^*}f$ for each complex fermion f,
are given by:
\beq
{Q(f) = {1\over 2}\alpha(f) + F(f)}
\label{qf}
\eeq
where $\alpha(f)$ is the boundary condition of the world--sheet fermion $f$
in the sector $\alpha$, and
$F_\alpha(f)$ is a fermion-number operator that takes the value $+1$ for
each mode of $f$, and the value $-1$ for each mode of $f^*$,
if $f$ is complex. For periodic fermions, which have
$\alpha(f)=1$, the vacuum
must be a spinor in order to represent the Clifford
algebra of the corresponding zero modes. For each periodic complex fermion
$f$, there are two degenerate vacua ${\vert +\rangle},{\vert -\rangle}$,
annihilated by the zero modes $f_0$ and
${{f_0}^*}$, respectively, and with fermion numbers  $F(f)= \pm 1$.

Realistic models in this free-fermionic formulation are generated by
a suitable choice of boundary-condition basis vectors for all world--sheet
fermions, which may be constructed in two stages. The first stage consists
of the NAHE set \cite{fsu5,nahe} of five boundary condition basis
vectors, $\{{{\bf 1},S,b_1,b_2,b_3}\}$.
After generalized GSO projections over the NAHE set, the residual gauge
group
is $SO(10)\times SO(6)^3\times E_8$ with $N=1$ space--time
supersymmetry~\footnote{The vector $S$ in this NAHE set is the
supersymmetry generator, and the superpartners of
the states from a given sector $\alpha$ are obtained from the sector
$S+\alpha$.}. The space--time vector bosons that generate the gauge group
arise from the Neveu--Schwarz sector and from the sector $1+b_1+b_2+b_3$.
The Neveu--Schwarz sector produces the generators of
$SO(10)\times SO(6)^3\times SO(16)$. The sector $1+b_1+b_2+b_3$
produces the spinorial 128 of $SO(16)$ and completes the hidden-sector
gauge group to $E_8$.
The vectors $b_1$, $b_2$ and $b_3$ correspond to the three twisted
sectors in the corresponding orbifold formulation, and produce
48 spinorial 16-dimensional representations of $SO(10)$, sixteen each from
the sectors $b_1$, $b_2$ and $b_3$.

The second stage of the basis construction consists of adding three
more basis vectors to the NAHE set, corresponding
to Wilson lines in the orbifold formulation, whose general
forms are constrained by string consistency conditions such as
modular invariance, as well as by space-time supersymmetry.
These three additional vectors are needed to reduce the number of
generations
to three, one from each of the sectors $b_1$, $b_2$ and $b_3$.
The details of the additional basis vectors distinguish between different
models
and determine their phenomenological properties.

The residual three generations constitute representations of the final
observable gauge group, which can be $SU(5)\times U(1)$~\cite{fsu5},
$SO(6)\times SO(4)$~\cite{alr} or $SU(3)\times SU(2)\times
U(1)^2$~\cite{fny,eu,lykken}.
In the former two cases, an additional pair of $16$ and $\overline{16}$
representations of $SO(10)$
is obtained from the two basis vectors that extend the NAHE set.
The electroweak Higgs multiplets are obtained from the Neveu--Schwarz
sector,
and from a sector that is a combination of the two basis vectors
which extend the NAHE set. This combination has the property that
$X_R\cdot X_R=4$, and produces states that transform solely
under the observable-sector symmetries.
Massless states from this sector are then obtained by
acting on the vacuum with fermionic oscillators with
frequency $1/2$. Details of the flavor symmetries differ
between models, but consist of at least three $U(1)$ symmetries
coming from the observable-sector $E_8$. Additional flavor $U(1)$
factors
arise from the complexification of real world--sheet fermions,
corresponding to the internal manifold in a bosonic formulation.
The models typically contain a hidden sector
in which the final gauge group is a subgroup of the hidden $E_8$,
and three matter representations in the vectorial $16$ of $SO(16)$,
which arise from the breaking of the hidden $E_8$ to
$SO(16)$~\footnote{In general, the
models also contain exotic massless states that arise
from the breaking of the non--Abelian $SO(10)$ symmetry at
the string level.}.

The cubic and higher-order terms in the superpotential are obtained
by evaluating the correlators
\beq
A_N\sim \langle V_1^fV_2^fV_3^b\cdots V_N\rangle,
\label{supterms}
\eeq
where $V_i^f$ $(V_i^b)$ are the fermionic (scalar) components
of the vertex operators, using the rules given in~\cite{kln}.
Generically, correlators of the form (\ref{supterms}) are of order
${\cal O} (g^{N-2})$, and hence of progressively higher orders
in the weak-coupling limit.
One of the $U(1)$ factors in the free-fermion models is anomalous,
and generates a Fayet--Ilioupolos term which breaks supersymmetry
at the Planck scale. The anomalous $U(1)$ is broken, and supersymmetry
is restored, by a non--trivial VEV for some scalar
field that is charged under the anomalous $U(1)$.
Since this field is in general also charged with respect
to the other anomaly-free $U(1)$ factors, some non-trivial
set of other fields must also get non--vanishing VEVs $\cal V$,
in order to ensure that the vacuum is supersymmetric.
Some of these fields will appear in the nonrenormalizable terms
(\ref{supterms}), leading to
effective operators of lower dimension. Their coefficients contain
factors of order ${\cal V} / M$, which may not be very small,
particularly in the context of the M-theory resolution of the
unification-scale problem.

This technology has previously been used to study the issue of proton
decay in the context of $SU(5) \times U(1)$ and $SU(3)\times
SU(2)\times U(1)^2$ type models~\cite{fsu5pdecay}.
Here we study this issue
using two specific free-fermion models~\cite{eu,cus} as case studies.
In both models, the color-triplet Higgs multiplets from the
Neveu--Schwarz
sectors are projected out by a superstring doublet--triplet
splitting mechanism, so that conventional GUT-scale dimension-five
operators are absent. Whilst
the model of~\cite{eu}
contains one pair of color-triplet Higgs fields from the
sector $b_1+b_2+\alpha+\beta$, in the model of~\cite{cus}
the Higgs color triplets from this sector are projected
out by the generalized GSO projections. The two models
also contain exotic color triplets from sectors
that arise from the $SO(10)$ Wilson-line breaking,
and carry lepton numbers $\pm1/2$.

Examining the superpotential terms in the first model~\cite{eu},
we find the following non-renormalizable terms at order $N=6$
\beq
Q_3Q_2Q_2L_3{\Phi_{45}}{\bar\Phi}_2^-~~~{\rm and}~~~
Q_3Q_1Q_1L_3{\Phi_{45}}{\Phi}_1^+
\label{n6terms}
\eeq
Thus, dangerous dimension-five operators arise in this model
if either of the sets of fields $\{{\Phi}_{45}~;~{\bar\Phi}_2^-\}$
or $\{{\Phi}_{45}~;~{\Phi}_1^+\}$ gets a VEV in the cancellation of
the anomalous $U(1)$ D--term equations.
Even if we can choose flat directions such that these
order $N=6$ terms are suppressed, higher-order terms
can still generate the dangerous operators.
If the suppression factor $\langle\phi\rangle/M\sim1/10$,
the dimension-five
terms are suppressed by $\sim10^{N-2}$ in each successive order.
The order $N=6$ terms would certainly lead
to proton decay at a rate that contradicts experiment,
and the same would be true of many higher-order operators in
the proposed M-theory context.

Next we turn to the model of~\cite{cus}, which
introduces a new feature. In this model, the observable-sector
gauge group formed by the gauge bosons from the Neveu--Schwarz
sector alone is
\beq
SU(3)\times SU(2)\times U(1)_C\times U(1)_L\times U(1)_{1,2,3,4,5,6}~.
\label{cusns}
\eeq
However, in this model there are
two additional gauge bosons from the vector
combination ${\bf1}+\alpha+2\gamma$~\cite{cus},
where the vector $2 \gamma$ has periodic boundary
conditions for the internal fermions $\{{\bar\psi}^{1,\cdots,5},
{\bar\eta}^1,{\bar\eta}^2,{\bar\eta}^3,{\bar\phi}^{1,\cdots,4}\}$
and anti--periodic boundary conditions for
all the remaining world--sheet fermions:
\beq
2\gamma=(0,\cdots,0\vert{\underbrace{1,\cdots,1}_{
{\bar\psi^{1,\cdots,5}},{\bar\eta^{1,2,3}},
{\bar\phi}^{1,\cdots,4}}},0,\cdots,0)~.
\label{vector2gamma}
\eeq
This model also has two new gauge generators, whose gauge bosons
are singlets
of the non--Abelian group, but carry $U(1)$ charges.
Referring to these two generators as $T^\pm$, we can define
the linear combination
\beq
T^3\equiv {1\over4}\left[U(1)_C+U(1)_L+U(1)_4+U(1)_5+
			U(1)_6+U(1)_7-U(1)_9\right]
\label{T3}
\eeq
such that the three generators $\{T^\pm,T_3\}$ together
form the enhanced symmetry group $SU(2)$.
Thus, the original observable symmetry
group (\ref{cusns}) has been enhanced
to
\beq
       SU(3)_C \times SU(2)_L \times SU(2)_{cust} \times
             U(1)_{C'} \times U(1)_L
           \times U(1)_{1,2,3} \times U(1)_{4',5',7''}
\label{enhancedgroup}
\eeq
and the remaining $U(1)$ combinations which are orthogonal to
$T^3$ are given by
\beqn
   U(1)_{C^\prime}&\equiv&{1\over3}\,U(1)_C-{1\over2}\,U(1)_{7} +{1\over
2}\,U(1)_9~\nonumber\\
   U(1)_{4^\prime}&\equiv&U(1)_4-U(1)_5\nonumber\\
   U(1)_{5^\prime}&\equiv&U(1)_4+U(1)_5-2\,U(1)_6\nonumber\\
   U(1)_{7^{\prime\prime}}&\equiv&U(1)_C-{5\over3}\biggl\lbrack
U(1)_4+U(1)_5+U(1)_6\biggr\rbrack
           +U(1)_{7} -U(1)_9~.
\label{cuscom}
\eeqn
The weak hypercharge can still be defined as the linear
combination $1/3U(1)_C+1/2U(1)_L$. However, as the $U(1)_C$
symmetry is now part of the extended $SU(2)$ gauge group,
$U(1)_C$ is given as a linear combination of the generators above
\beq
    {1\over3}\,U(1)_{C}~=~{2\over5}\biggl\lbrace U(1)_{C^\prime}+
     {5\over{16}}\,\biggl\lbrack T^3+{3\over5}\,U_{7^{\prime\prime}}
\biggr\rbrack
        \biggr\rbrace~.
\label{U1Cin274}
\eeq
Since the weak hypercharge is not orthogonal
to the enhanced $SU(2)$ symmetry, it is convenient to define a new linear
combination of the $U(1)$ factors:
\beqn
        U(1)_{Y'} &\equiv& U(1)_Y - {1\over 8}\,T^3 \nonumber\\
        &=& {1\over 2}\,U(1)_L + {5\over{24}}\,U(1)_C \nonumber\\
         &&~~~~~-{1\over8}\,\biggl\lbrack
U(1)_4+U(1)_5+U(1)_6+U(1)_7-U(1)_9\biggr\rbrack~,
\label{U1pin274}
\eeqn
so that the weak hypercharge is expressed in terms of $U(1)_{Y'}$ as
\beq
        U(1)_{Y} = U(1)_{Y'} + {1\over 2}\,T^3 ~~~~\Longrightarrow~~~~
           Q_{\rm e.m.} = T^3_L + Y = T^3_L + Y' +
{1\over 2}\,T^3_{cust}~.\label{Qemin274}
\eeq
The final observable-sector gauge group is therefore
\beq
       SU(3)_C \times SU(2)_L \times SU(2)_{cust}\times U(1)_{Y'} ~\times
        ~\biggl\lbrace ~{\rm seven~other~}U(1){\rm ~factors}~\biggr\rbrace~.
\label{finalgroup}
\eeq
The remaining seven $U(1)$ factors must be chosen
as linear combinations of the previous $U(1)$ factors so as to be orthogonal
to the each of the other factors in (\ref{finalgroup}).

Up to $U(1)$ charges, the massless spectrum from the Neveu--Schwarz
sector and the sector $b_1+b_2+\alpha+\beta$ is the same as in the
previous model. However,
because of the enhanced symmetry, the spectrum from the sectors $b_j$
is modified. The sectors $b_j\oplus{\bf1}+\alpha+2\gamma$.
produce the three light generations, one from each of the
sectors $b_j$ $(j=1,2,3)$, as before. However, the gauge
enhancement noted above has the important corollary that
only the leptons, $\{L,e_L^c,N_L^c\}$, transform as doublets of the
enhanced $SU(2)_{cust}$ gauge group, whilst the quarks,
$\{Q,u^c_L,d_L^c\}$,
are $SU(2)_{cust}$ singlets. Therefore, terms of the form $QQQL$ are
not invariant under the enhanced $SU(2)_{cust}$ gauge group. Furthermore,
such terms are forbidden to all orders of non-renormalizable terms,
as can be verified by an explicit computerized search.
Even if we break the custodial $SU(2)_{cust}$ by, \eg, the VEV of
the right--handed neutrino and its complex conjugate,
the higher-order terms will not be invariant under the
combined symmetries $SU(2)_{cust}$ and $U(1)_L$. A
computerized search for all possible operators
that might lead to proton decay confirms that
such terms do not arise at any order in this model.

This model therefore
provides an example how the proton decay problem
may be resolved in a robust way: even if the string
unification scale is lowered to the minimal supersymmetric GUT scale, as
proposed
in the M--theory strong-coupling solution to the string
gauge-coupling unification problem, such a model can
evade the proton decay constraints to all orders in
perturbation theory. However, we recognize that this approach does not
encompass strictly non-perturbative string effects which may
appear in a direct M- or F-theory construction. On the other
hand, we also note
that enhanced gauge symmetries appear in many M- and F-theory
constructions~\cite{enhanced}, and may play a role analogous to that
played
by the enhancement in the above model.

Such enhanced symmetries arise frequently
in the free-fermion models,
whenever there is a combination of
the basis vectors which extends the NAHE set:
\beq
X=n_\alpha\alpha+n_\beta\beta+n_\gamma\gamma
\label{combi}
\eeq
for which $X_L\cdot X_L=0$ and $X_R\cdot X_R\ne0$. Such a
combination may produce additional space--time vector
bosons, depending on the choice of generalized GSO phases.
For example, in the flipped $SU(5)$ model of~\cite{price},
in addition to the gauge bosons from the Neveu--Schwarz sector
and the sector $I={\bf1}+b_1+b_2+b_3$,
additional space--time vector bosons are obtained from the
sectors $b_1+b_4\pm\alpha\oplus I$.
In this case, the hidden-sector $SU(4)$ gauge group, arising from
the gauge bosons of the NS$\oplus I$ sectors, is
enhanced to $SU(5)$. This particular enhancement does not
modify the observable gauge sector, and does nothing to
forbid dangerous higher-order operators.
However, other, more interesting, cases may exist.

The type of
enhancement depends not only on the boundary-condition basis vectors,
but also on the discrete choices of
GSO phases. For example, in the model of~\cite{eu},
the combination
of basis vectors $X=b_1+b_2+b_3+\alpha+\beta+\gamma+(I)$
has $X_L\cdot X_L=0$, and thus may give rise to additional
space--time vector bosons. All the extra space--time
vector bosons are projected out by the choice of generalized
GSO projection coefficients.
However, with the modified GSO phases
\beq
 c\left(\matrix{1\cr\gamma\cr}\right)\rightarrow
-c\left(\matrix{1\cr\gamma\cr}\right),
 c\left(\matrix{\alpha\cr\beta\cr}\right)\rightarrow
-c\left(\matrix{\alpha\cr\beta\cr}\right) ~\hbox{and}~
 c\left(\matrix{\gamma\cr\beta\cr}\right)\rightarrow
-c\left(\matrix{\gamma\cr\beta\cr}\right),
\label{eumodi1}
\eeq
additional space--time
vector bosons are obtained from the sector
$b_1+b_2+b_3+\alpha+\beta+\gamma+(I)$. In addition,
the sector $b_1+b_2+b_3+\alpha+\beta+\gamma+(I)$
produces the representations $3_1+3_{-1}$ of $SU(3)_H$, where
one of the $U(1)$ combinations is the $U(1)$ in the decomposition
of $SU(4)$ under $SU(3)\times U(1)$.
In this case, the hidden-sector $SU(3)_H$ gauge group is extended to
$SU(4)_H$. If instead we take the modified phases
\beq
 c\left(\matrix{1\cr\gamma\cr}\right)\rightarrow
-c\left(\matrix{1\cr\gamma\cr}\right),
 c\left(\matrix{\gamma\cr\alpha\cr}\right)\rightarrow
-c\left(\matrix{\gamma\cr\alpha\cr}\right),
\label{eumodi2}
\eeq
then the sector ${\bf1}+\alpha+\beta+\gamma$ produces
two additional space--time vector bosons which enhances
one of the $U(1)$ factors to $SU(2)$. These examples further
illustrate the point that this type of enhancement is common in
realistic free-fermion models, which makes it interesting
to explore further in connection with the
proton stability problem.

We now explore the possibility of
a connection between the type of models discussed
above and the vacua of M (and F) theory.
At present, a direct connection between
known features of the non-perturbative formulation of M theory and
the above realistic free-fermion models, with a full basis consisting of
eight vectors, is not yet possible. However, it is nevertheless
possible to make observations suggesting
a possible connection of these models to the type of M-- (and F--)theory
compactifications which  have been discussed in the
literature~\cite{z2mf}.

We start by studying in more detail the geometric interpretation
of the five-basis-vector NAHE set, $\{{\bf1},S,b_1,b_2,b_3\}$
that underlies the realistic
free fermionic models.
This set corresponds
to a $Z_2\times Z_2$ orbifold compactification
of the weakly-coupled ten-dimensional heterotic string, and the basis
vectors $b_1$, $b_2$ and $b_3$ correspond to the three twisted
sectors of these orbifold models. To see this correspondence,
we add to the NAHE set the basis vector
\beq
X=(0,\cdots,0\vert{\underbrace{1,\cdots,1}_{{\bar\psi^{1,\cdots,5}},
{\bar\eta^{1,2,3}}}},0,\cdots,0)~.
\label{vectorx}
\eeq
with the following choice of generalized GSO projection coefficients:
\begin{equation}
      C\left( \matrix{X\cr \b_j\cr}\right)~=~
      -C\left( \matrix{X\cr S\cr}\right) ~=~
      C\left( \matrix{X\cr \bone \cr}\right) ~= ~ +1~.
\label{Xphases}
\end{equation}
This set of basis vectors produces models with an
$SO(4)^3\times E_6\times U(1)^2\times E_8$ gauge group
and $N=1$ space--time supersymmetry. The matter fields
include 24 generations in 27 representations of
$E_6$, eight from each of the sectors $b_1\oplus b_1+X$,
$b_2\oplus b_2+X$ and $b_3\oplus b_3+X$.
Three additional 27 and $\overline{27}$ pairs are obtained
from the Neveu--Schwarz $\oplus~X$ sector.

The subset of basis vectors
\beq
\{{\bf1},S,X,I={\bf1}+b_1+b_2+b_3\}
\label{neq4set}
\eeq
generates a toroidally-compactified model with $N=4$ space--time
supersymmetry and $SO(12)\times E_8\times E_8$ gauge group.
The same model is obtained in the geometric (bosonic) language
by constructing the background fields which produce
the $SO(12)$ Narain lattice~\cite{Narain,foc}, taking the metric
of the six-dimensional compactified manifold
to be the Cartan matrix of $SO(12)$:
\beq
g_{ij}=\left(\matrix{~2&-1& ~0& ~0& ~0& ~0\cr%
-1& ~2&-1& ~0& ~0& ~0\cr~0&-1& ~2&-1& ~0& ~0\cr~0& ~0&-1
& ~2&-1&-1\cr ~0& ~0& ~0&-1& ~2& ~0\cr ~0& ~0& ~0&-1& ~0& ~2\cr}\right)
\label{gso12}
\eeq
and the antisymmetric tensor
\beq
b_{ij}=\cases{
g_{ij}&;\ $i>j$,\cr
0&;\ $i=j$,\cr
-g_{ij}&;\ $i<j$.\cr}
\label{bso12}
\eeq
When all the radii of the six-dimensional compactified
manifold are fixed at $R_I=\sqrt2$, it is easily seen that the
left-- and right--moving momenta
\beq
P^I_{R,L}=[m_i-{1\over2}(B_{ij}{\pm}G_{ij})n_j]{e_i^I}^*
\label{lrmomenta}
\eeq
reproduce all the massless root vectors in the lattice of
$SO(12)$,
where in (\ref{lrmomenta}) the $e^i=\{e_i^I\}$ are six linearly-independent
vectors normalized: $(e_i)^2=2$.
The ${e_i^I}^*$ are dual to the $e_i$, and
$e_i^*\cdot e_j=\delta_{ij}$.

Adding the two basis vectors $b_1$ and $b_2$ to the set
(\ref{neq4set}) corresponds to the $Z_2\times Z_2$
orbifold model with standard embedding.
The fermionic boundary conditions are translated
in the bosonic language to twists on the internal dimensions
and shifts on the gauge degrees of freedom.
Starting from the Narain model with $SO(12)\times E_8\times E_8$
symmetry~\cite{Narain}, and applying the $Z_2\times Z_2$ twisting on the
internal
coordinates, we then obtain the orbifold model with $SO(4)^3\times
E_6\times U(1)^2\times E_8$ gauge symmetry. There are sixteen fixed
points in each twisted sector, yielding the 24 generations from the
three twisted sectors mentioned above. The three additional pairs of $27$
and $\overline{27}$
are obtained from the untwisted sector. This
orbifold model exactly corresponds to the free-fermion model
with the six-dimensional basis set
$\{{\bf1},S,X,I={\bf1}+b_1+b_2+b_3,b_1,b_2\}$.
The Euler characteristic of this model is 48 with $h_{11}=27$ and
$h_{21}=3$.

This $Z_2\times Z_2$ orbifold, corresponding
to the extended NAHE set at the core of the realistic
free fermionic models,
differs from the one which has usually been
examined in the literature~\cite{z2mf}.
In that orbifold model, the Narain
lattice is $SO(4)^3$, yielding a $Z_2\times Z_2$ orbifold model
with Euler characteristic equal to 96, or 48 generations,
and $h_{11}=51$, $h_{21}=3$.

In more realistic free-fermion models, the vector $X$
is replaced by the vector $2\gamma$ (\ref{vector2gamma}).
This modification has the consequence of producing a
toroidally-compactified model with $N=4$ space--time supersymmetry and
gauge group $SO(12)\times SO(16)\times SO(16)$.
The $Z_2\times Z_2$ twisting breaks the gauge symmetry to
$SO(4)^3\times SO(10)\times U(1)^3\times SO(16)$.
The orbifold twisting still yields a model with 24 generations,
eight from each twisted sector,
but now the generations are in the chiral 16 representation
of $SO(10)$, rather than in the 27 of $E_6$. The same model can
be realized with the set
$\{{\bf1},S,X,I={\bf1}+b_1+b_2+b_3,b_1,b_2\}$,
projecting out the $16\oplus{\overline{16}}$
from the sector $X$ by taking
\beq
c{X\choose I}\rightarrow -c{X\choose I}.
\label{changec}
\eeq
This choice also projects out the massless vector bosons in the
128 of $SO(16)$ in the hidden-sector $E_8$ gauge group, thereby
breaking the $E_6\times E_8$ symmetry to
$SO(10)\times U(1)\times SO(16)$.
This analysis confirms that the $Z_2\times Z_2$ orbifold on the
$SO(12)$ Narain lattice is indeed at the core of the
realistic free fermionic models.

We can now examine whether some connection with
M- (and F-)theory compactifications can be contemplated,
in view of the extensive literature on
$Z_2\times Z_2$ orientifolds of M and F theory \cite{z2mf}.
In particular, these interesting papers have examined
in detail the $Z_2\times Z_2$
orbifold model with $h_{11}=51$ and $h_{21}=3$.
This model is precisely the $Z_2\times Z_2$ orbifold model
obtained by twisting the $SO(4)^3$ Narain lattice.
In this compactification, the six-dimensional compactified
space is the direct product of three simple two-tori, {\it i.e.},
$(T_2)^3$.
This model has been investigated extensively in~\cite{z2mf}
in connection with M-- and F--theory compactifications
on special classes
of Calabi--Yau threefolds that have been analyzed by
Voisin~\cite{voisin} and Borcea~\cite{borcea}.
They have been further classified by
Nikulin \cite{nikulin} in terms of three invariants $(r,a,\delta)$,
in terms of which $(h_{1,1},h_{2,1})=
(5+3r-2a,65-3r-2a)$. Within this framework, the
$(h_{1,1},h_{2,1})=(51,3)$ $Z_2\times
Z_2$ orbifold model coincides with the Voisin--Borcea model
with $(r,a,\delta)=(18,4,0)$.
The dual
relations between compactifications of this manifold on
M--, F--theory compactifications and type IIB orientifolds
have been demonstrated in~\cite{z2mf}.

The task ahead is clear. Naively, the (27,3) $Z_2\times Z_2$
orbifold would correspond to a Voisin--Borcea model with
$(r,a,\delta)=(14,10,0)$. However, such a Voisin--Borcea model
is not (yet) known to exist. The problem is that in
the Voisin--Borcea models the factorization
of the six-dimensional manifold as a product of three disjoint
manifolds is essential. However, our $(27,3)$ $Z_2\times Z_2$
model, being an orbifold of a $SO(12)$ lattice, is an intrinsically
$T^6$ manifold, and, as such, factorization \`a la Voisin-Borcea is not
possible. Therefore, the first step
in trying to connect realistic free-fermion
models to M-- and F--theory compactifications is to construct
the Calabi--Yau threefolds which correspond to the $Z_2\times Z_2$
orbifold on the $SO(12)$ lattice. Although the relevant manifolds,
or their Landau--Ginzburg potential realizations, are not yet known
(at least not to us), we believe that they are not fundamentally different
in nature from, or
intrinsically more difficult than, the corresponding
orientifolds related to the
$Z_2\times Z_2$ orbifold on $SO(4)^3$ lattice.

We now recap the current status of the effort to connect
M-- and F--theory compactifications to relevant phenomenological
data. On the one hand, we have the appealing
free-fermion models, in which one can address in detail many relevant
phenomenological questions, and
which provide promising candidates for
a realistic superstring model.
We have examined in detail in this paper the issue of proton stability,
and shown how enhanced gauge symmetries which prevent fast
proton can arise. Thus, we have exhibited a robust solution to
the problem of the proton lifetime, which is of crucial relevance
for M--theory compactifications.

As a step towards the elevation of these ideas to M and F theory,
we have discussed the orbifold correspondences
of these models. At the core of the realistic free-fermion
models there is a $Z_2\times Z_2$ orbifold on an $SO(12)$
lattice with $(h_{1,1},h_{2,1})=(27,3)$. This is not
the standard $Z_2\times Z_2$ orbifold that has been discussed
extensively in the literature, the more familiar one being that
with $(h_{1,1},h_{2,1})=(51,3)$. Nevertheless, the existence
of duality relations between the $(51,3)$ $Z_2\times Z_2$ orbifold
and M-- and F--theory compactifications leads us to expect the
existence of similar relations for the $(27,3)$ $Z_2\times Z_2$
orbifold. If the relevant Calabi-Yau threefold or its
Landau--Ginzburg realization can indeed be found,
the connection of M and F theory to relevant low-energy
data would have made a major step forward, particularly with regard
to proton stability. Such progress is not out of sight.


\bigskip
\medskip
\leftline{\large\bf Acknowledgments}
\medskip

We are pleased to thank Julie Blum and Cumrun Vafa for discussions.
This work was supported in part by the Department of Energy
under Grants No.\ DE-FG-05-86-ER-40272 and DE-FG05-91-GR-40633.


\vfill\eject

\bibliographystyle{unsrt}

\end{document}